\documentclass[usegraphicx,usenatbib,onecolumn,doublespace]{mn2e}
\usepackage{amsmath}
\usepackage{amssymb}
\usepackage{bm}

\begin{document}

\title{The effect of primordial black holes on 21cm fluctuations}

\author[Tashiro, H. et al.]
{Hiroyuki Tashiro$^1$\
 and Naoshi Sugiyama$^{2,3}$\\
  $^1$Physics Department, Arizona State University, Tempe, AZ 85287, USA\\
  $^2$Department of Physics and Astrophysics \& Kobayashi-Maskawa
  Institute for the Origin of Particles and the Universe,\\
  Nagoya University, 
  Chikusa, Nagoya 464-8602, Japan\\
  $^3$Institute for the Physics and Mathematics of the Universe (IPMU),
The University of Tokyo, Kashiwa, Chiba, 277-8568, Japan
  }

\date{\today}

\maketitle

\begin{abstract}

 The 21~cm signal produced by non-evaporating primordial black holes
(PBHs) is investigated.  X-ray photons emitted by accretion of matter onto a PBH
ionize and heat the intergalactic medium (IGM) gas near the PBH. Using a
simple analytic model, we show that this X-ray heating can produce an
observable differential 21~cm brightness temperature.  The region of
 the observable 21~cm brightness temperature
can extend to 1--10 Mpc comoving distance from a PBH depending on
 the PBH mass.  The angular power
spectrum of 21~cm fluctuations due to PBHs is also calculated. The peak
position of the angular spectrum depends on PBH mass, while the
amplitude is independent of PBH mass.  Comparing this angular power spectrum
 with the angular power spectrum
caused by primordial density fluctuations, it is found that both of them
become comparable if $\Omega_{{\rm PBH}} = 10^{-11} (M/10^{3}~%
M_\odot)^{-0.2}$ at $z=30$ and $10^{-12} (M/10^{3}~ M_\odot)^{-0.2}$ at
$z=20$ for the PBH mass from $10 ~ M_\odot $ to $10^8~ M_\odot $.  Finally
we find that the Square Kilometer Array can detect the signal due to
PBHs up to $\Omega_{\rm PBH}=10^{-5} (M/10^{3}~ M_\odot)^{-0.2}$ at
$z=30$ and $10^{-7} (M/10^{3}~ M_\odot)^{-0.2}$ at $z=20$ for PBHs with
mass from $10^2 ~ M_\odot $ to $10^8~ M_\odot $.

\end{abstract}

\begin{keywords}
cosmology: theory -- large-scale structure of universe

\end{keywords}

\maketitle

\section{Introduction}

primordial black holes (PBHs) could have formed in the early
Universe \citep{1974MNRAS.168..399C,1975ApJ...201....1C}.  Although there is no
direct evidence of their existence of PBHs, PBHs are attracting attention
as a way of constraining physics in the early Universe.  In
particular, one of the main generation mechanisms of PBHs is the gravitational
collapse of an overdense region at the horizon scale when the
amplitude of the overdensity exceeds a critical threshold.
Therefore the resultant mass function and
the abundance of PBHs depend on the amplitude of primordial density
fluctuations at the horizon-crossing epoch \citep{2004PhRvD..70d1502G}.
The PBH abundance is expected to be a probe of 
primordial density fluctuations on small scales, which cannot be accessed
by Cosmic Microwave Background or large-scale-structure observations.

Constraints on the abundance of PBHs have been extensively studied 
and continue to be updated \citep{2010PhRvD..81j4019C}.  PBHs with mass
less than $10^{15}~$g have evaporated by the present epoch because the
evaporation time scale by Hawking radiation is less than the Hubble time scale today
\citep{1974Natur.248...30H}.  However evaporation of 
PBHs generates additional entropy in the Universe after inflation
\citep{
1976JETPL..24..571Z},
affects big bang nucleosynthesis
\citep{
1978SvAL....4..185V, 
1978SvA....22..138V,1978PThPh..59.1012M,
1977SvAL....3..110Z,1980MNRAS.193..593L}
and distorts the CMB blackbody spectrum \citep{2008PhRvD..78b3004T}.
PBH evaporation may also produce the observable gamma-ray background \citep{1976ApJ...206....1P,1991ApJ...371..447M}.
According to measurements of these cosmological
phenomena, there are strong constraints on the abundance of PBHs
with mass less than $10^{15}~$g. 

PBHs with mass larger than $10^{15}~$g survive in the present Universe.
One of the constraints on such PBHs can be set from the
fact that the current density parameter of PBHs, $\Omega_{\rm PBH}$,
cannot exceed the cold dark matter density parameter observed at the
present epoch,
$\Omega_{\rm C}$.
Conventionally, the constraint on PBH 
abundance is given
by $\beta(M)$ which is the fraction of regions of mass $M$
collapsing into PBHs
at the formation epoch \citep{1975ApJ...201....1C}.   
The constraint on the density parameter of PBHs today, $\Omega_{{\rm
PBH}}<\Omega_{\rm C}$, implies
$\beta < 2 \times 10^{-18} (M/10^{15}{\rm g})^{1/2}$
from WMAP 7-year data, i.e., $\Omega_{\rm C} =0.22$ \citep{2011ApJS..192...18K}.  
Microlensing observations also constrain the abundance of
non-evaporating PBHs \citep{2001ApJ...550L.169A}.
\citet{2008ApJ...680..829R}
have obtained the constraint on PBHs with mass larger than $0.1~{\rm M}_\odot$,
by investigating the effects of such PBHs on cosmic reionization and CMB
temperature anisotropies.
Future gravitational wave observations are also expected to provide a
probe of the massive PBH abundance \citep{1999PhRvD..60h3512I,2003PhRvL..91b1101I}.

In this paper, we evaluate the 21~cm brightness temperature produced by
PBHs and study the potential of 21~cm observations to give a constraint
on the abundance of PBHs.  \citet{2008arXiv0805.1531M} have
investigated the signatures of evaporating PBHs in 21~cm brightness
temperature.  Accordingly, they have concentrated on PBHs whose mass range
is $5 \times 10^{13}~{\rm g} \lesssim M_{\rm PBH} \lesssim 10^{17}~{\rm g} $. On the
contrary, here, we focus on non-evaporating PBHs with mass much larger
than $10^{15}~$g.
After the epoch of matter-radiation equality, gas and matter can accrete onto PBHs.
It has been shown that PBHs with large mass
could produce X-ray and UV photons through the accretion of matter onto PBHs
and these photons heat up and
ionize intergalactic medium (IGM)~\citep{1981MNRAS.194..639C,1995ApJ...438...40G,2001ApJ...561..496M,2008ApJ...680..829R}.
Therefore, the heated and ionized IGM gas may produce an observable deviation of
the 21~cm brightness temperature from the background even before the
birth of the
first stars and galaxies~($z>30$).
Our aim in this
paper is to evaluate this deviation and to discuss the potential of 
21~cm observations to constrain the non-evaporating PBH abundance.

The paper is organized as follows. In section 2, using a simple model of
X-ray photon flux due to the accretion onto a PBH, we evaluate the
ionization and temperature profile near a PBH. In section 3, we calculate
the spin temperature and the brightness temperature induced by a PBH.
In section 4, the angular power spectrum of 21~cm fluctuations due to
PBHs are evaluated. Section 5 is devoted to the
conclusion.  
Throughout this paper, we use parameters for a flat $\Lambda$CDM model: 
$h=0.7$ $(H_0=h \times 100 ~ {\rm km /s /Mpc})$, $\Omega_{\rm B}=0.05$ and 
$\Omega_{\rm M}=0.26$. These parameters are consistent with  
WMAP results \citep{2011ApJS..192...18K}.

\section{Ionization and Heating of IGM by a PBH}

After the epoch of matter-radiation equality, matter can accrete onto a
non-evaporating PBH, whose mass is  larger than $10^{15}\rm g$.
Therefore, the accretion disk of a non-evaporating PBH can be a source of X-ray
photons before first stars and galaxies form ($z>30$).
For example,
the non-evaporating PBHs are one of the candidates for super massive
black hole seeds.
PBHs whose masses exceed $10^5~M_\odot$ cannot be
directly formed through gravitational collapse since the time scale of
the collapse becomes longer than cosmological time.   Accordingly
only the accretion after the formation, which induces X-ray photon
emission,  makes possible to 
form such massive black holes
\citep{2004PhRvD..70f4015D,2005APh....23..265K}.  
It is difficult to theoretically predict the accurate X-ray photon spectrum from PBHs, because
the X-ray spectrum depends on the detailed condition of the accretion
and the environment of PBHs such as the amount of neutral hydrogen.
Hence, for simplicity, we assume that the PBH accretion powers a
miniquasar with a power-law spectrum of X-ray photons, according to \cite{2005MNRAS.363.1069K},
\begin{equation}
 F(E) = {\cal A} E^{-1}~s^{-1}  ,
\end{equation}
where $\cal A $ is set to correspond to the tenth of the Eddington luminosity.  
Here we consider the range of the photon energy E from 200~eV to 100~keV
since we assume that emitted photons whose energy are lower than 200~eV are
immediately absorbed by the surrounding gas of the PBH.
Following \citet{2007MNRAS.375.1269Z},
we evaluate the ionization and heating of IGM due to massive PBHs in this section.

The number density of photons per unit time per unit area at distance
$r$ from the source is given by
\begin{equation} 
{\cal N}(E;r)
 = e^{-\tau(E;r)} \frac{{\cal A}}{\left(4 \pi
r^2\right)}
E^{-1}\mathrm{cm^{-2} s^{-1}},
\label{eq:flux_r}
\end{equation}
\begin{equation}
\tau(E;r)= \int_0^r n_H x_{H} \sigma(E) dr ,
\label{eq:flux_tau}
\end{equation}
where $x_{H}$ is the hydrogen neutral fraction, $n_H$ is
the mean number density of hydrogen at a redshift $z$ and
$\sigma(E)$ is the absorption cross-section per hydrogen atom.
In order to take into
 account the contribution from helium atoms as well as hydrogen atoms, 
we adopt the fitting formula by \citet{1989ApJ...344..551Z},
\begin{equation}
 \sigma = 4.25 \times 10^{-21} \left(   \frac{E}{250~ {\rm eV}}
\right)^{-p} ~{\rm cm^2},
 \quad p= 
  \begin{cases}
     2.65  \quad {\rm for}\ E< 250\ {\rm eV}, \\
         3.30  \quad {\rm for}\ E>250\ {\rm eV}. 
  \end{cases}
\label{eq:sigma}
\end{equation}
Using  the function ${\cal N}(E;r)$,
we can write the ionization rate per hydrogen atom at the
distance $r$ from the source as
\begin{equation}
 \Gamma (r) = \int_{E_0}^{\infty}\sigma(E) {\cal N}(E;r)
  \left( 1 + \frac{E}{E_0}~\phi(E,x_{e})\right)\frac{dE}{E},
\end{equation}
where
 the term, $(E/E_0)   \phi(E,x_e) $, is introduced  to consider 
the secondary ionization due to the photoelectrons
produced by energetic photons ($E > 100 ~\rm eV$).
We apply the fitting formula of $\phi$ by \citet{2004ApJ...613..646D}
for the low energy region $E<0.5~$keV and by \citet{1985ApJ...298..268S} 
for the high energy region $E>0.5~$keV.

\begin{figure}
 \begin{tabular}{cc}
 \begin{minipage}{0.5\hsize}
  \begin{center}
   \includegraphics[width=75mm]{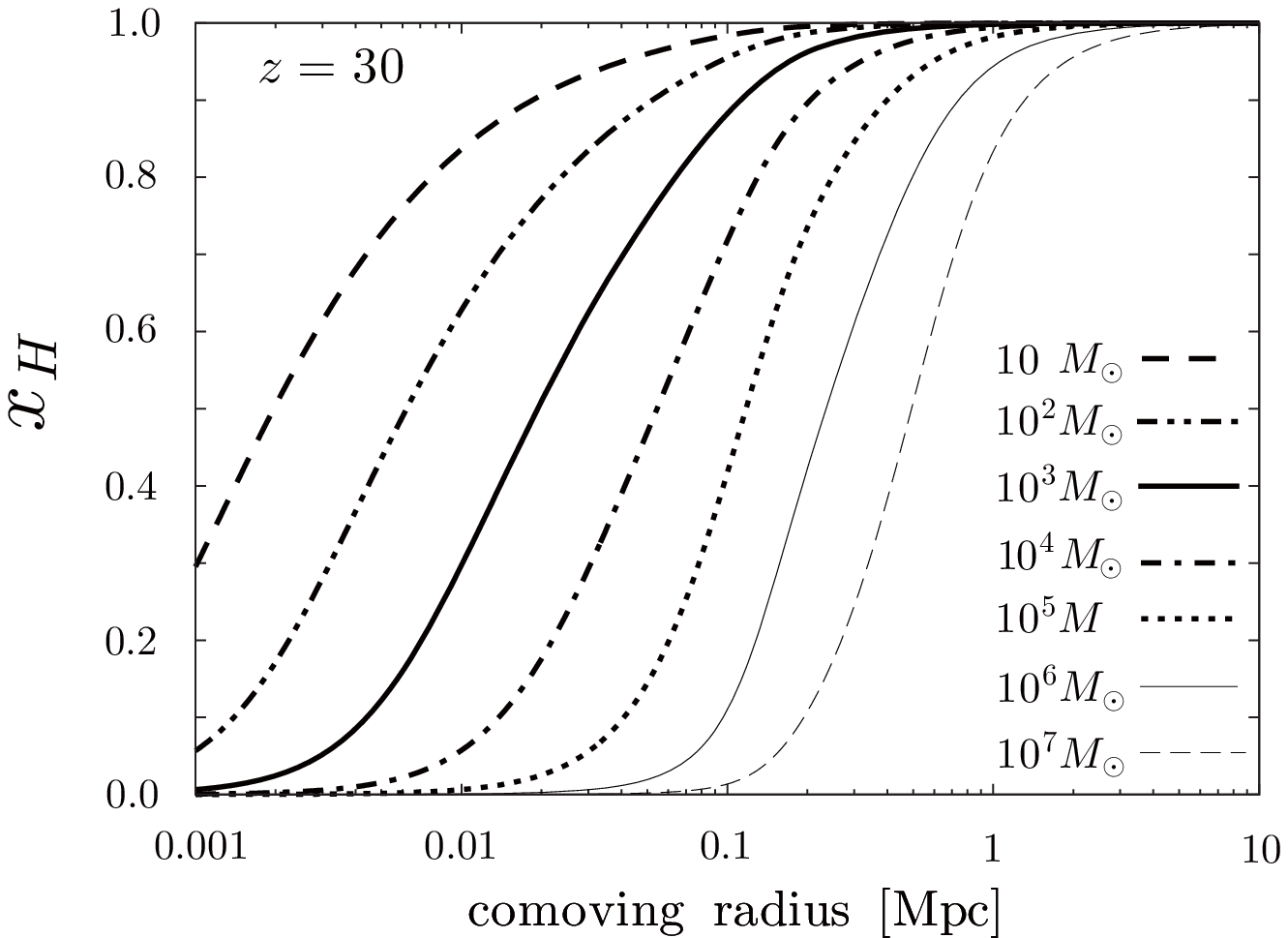}
  \end{center}
 \end{minipage}
 \begin{minipage}{0.5\hsize}
  \begin{center}
  \includegraphics[width=75mm]{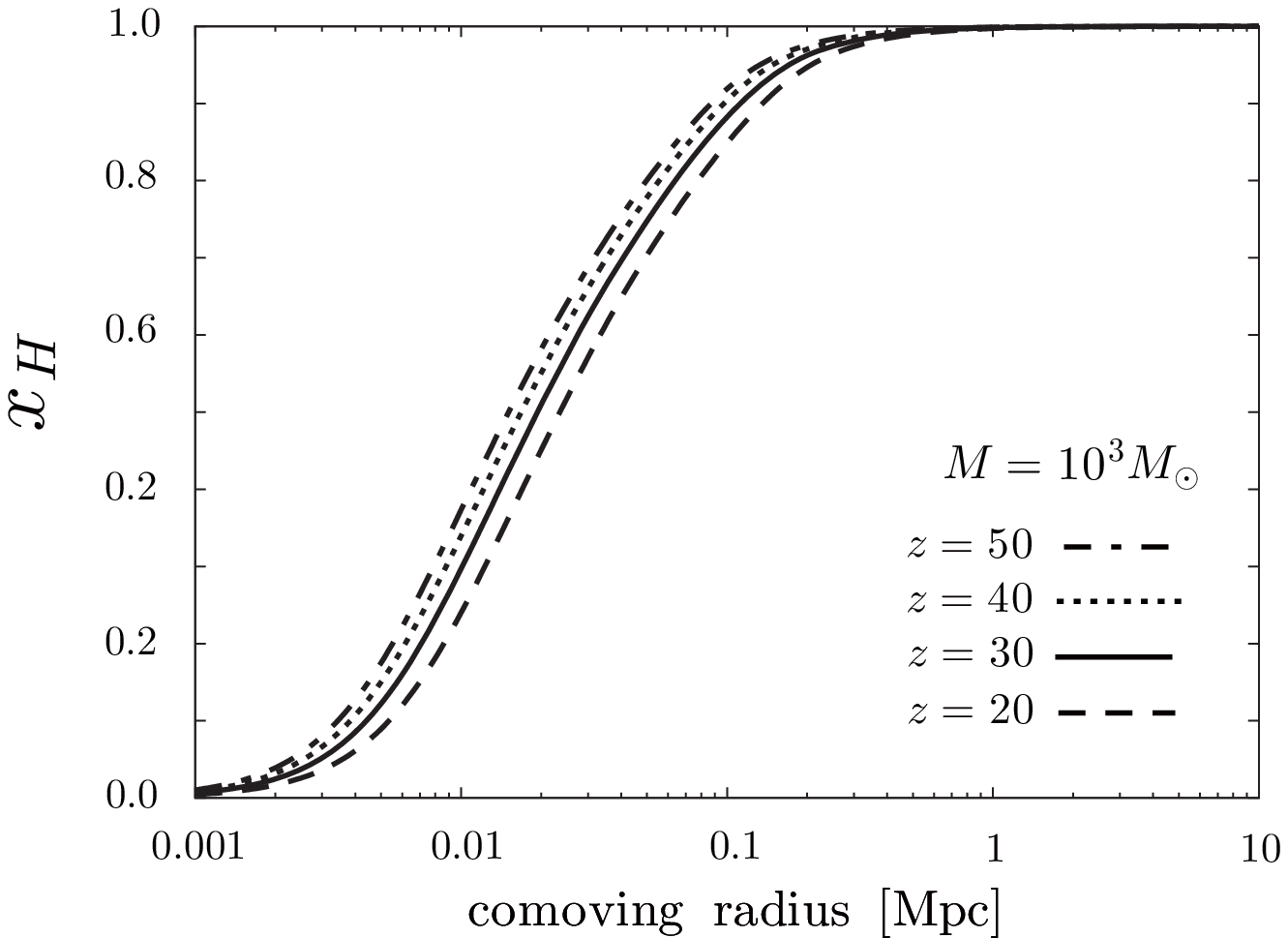}
  \end{center}
 \end{minipage}
\end{tabular}
\caption{The neutral fraction of hydrogen as a function of comoving
 distance from the source.  The left panel shows the
 dependence on mass at $z=30$. The lines represents the neutral fraction
 with masses ranging from $10~M_\odot$ to $10^7~M_\odot$ from left to right.
 The right panel shows the dependence on
 the redshift for a PBH with $M=10^3 ~M_\odot$. The dotted-dashed, dotted,
 solid and dashed lines represent the neutral fraction at $z=50$,
 $z=40$, $z=30$ and $z=20$, respectively.
 }
\label{fig:xh}
\end{figure}

The neutral fraction of hydrogen is obtained by solving the equation of
the ionization-recombination equilibrium 
\begin{equation}
\alpha_{H } n_H^2 (1-x_{H })^2= \Gamma(r)~n_H x_{H },
\label{eq:ionization_eq}
\end{equation}
where $\alpha_{H}$ is the recombination cross-section;
$\alpha_{H} =2.6 \times 10^{-13} 
\mathrm{cm^3 s^{-1}}$,

Solving Eq.~(\ref{eq:ionization_eq}), we show the neutral fraction of
hydrogen for different masses and for different redshifts in the left
and right panels of Fig.~\ref{fig:xh}, respectively. 
Here we assume that the density around a PBH is the same as the mean
density of the Universe. The comoving radius of the
ionization sphere is made large by a massive PBH. Increasing the mass of
the PBH means
increasing the number of the ionization photons, because we assume that the flux of the
ionization photons is proportional to a tenth of the Eddington luminosity.
As the Universe evolves,  the comoving radius
of the ionization sphere increases slowly.

Next, we evaluate the kinetic temperature of the IGM around a PBH.
The heating rate per unit volume per unit time at a distance $r$
from the source is obtained by considering the photons absorbed by
the IGM at $r$,
\begin{equation}
{\cal H}(r) = f n_{H} x_{H}(r)\int_{E_0}^{\infty}\sigma(E) {\cal N}(E;r)
dE,
\label{eq:heat}
\end{equation}
where $f$ is the fraction of the photon energy absorbed through the
collisional excitations of the IGM.
\citet{1985ApJ...298..268S} provided a simple fitting formula
$f = C \left[ 1- \left( 1- x^a\right)^b \right]$, where
$C=0.9771$, $a=0.2663,$ $b=1.3163$ and $x$ is the ionized
fraction $x=1-x_{H}$.

The kinetic temperature of the IGM at a distance $r$, $T_k(r)$, is determined by the balance between the
heating and the Compton cooling due to CMB photons,
\begin{equation}
{\cal H}(r) = \frac{8 \sigma_T}{3 m_e}
 T_{\gamma} ^4 (1-x_H) (T_k(r) - T_{\gamma}) + 2 H T_k(r), 
\end{equation}
where $\sigma_T$ is the cross section for the Compton scattering and
$T_\gamma$ is CMB temperature.   Here we also take into account
the cooling by the expansion of the Universe.

\begin{figure}
 \begin{tabular}{cc}
 \begin{minipage}{0.5\hsize}
  \begin{center}
   \includegraphics[width=75mm]{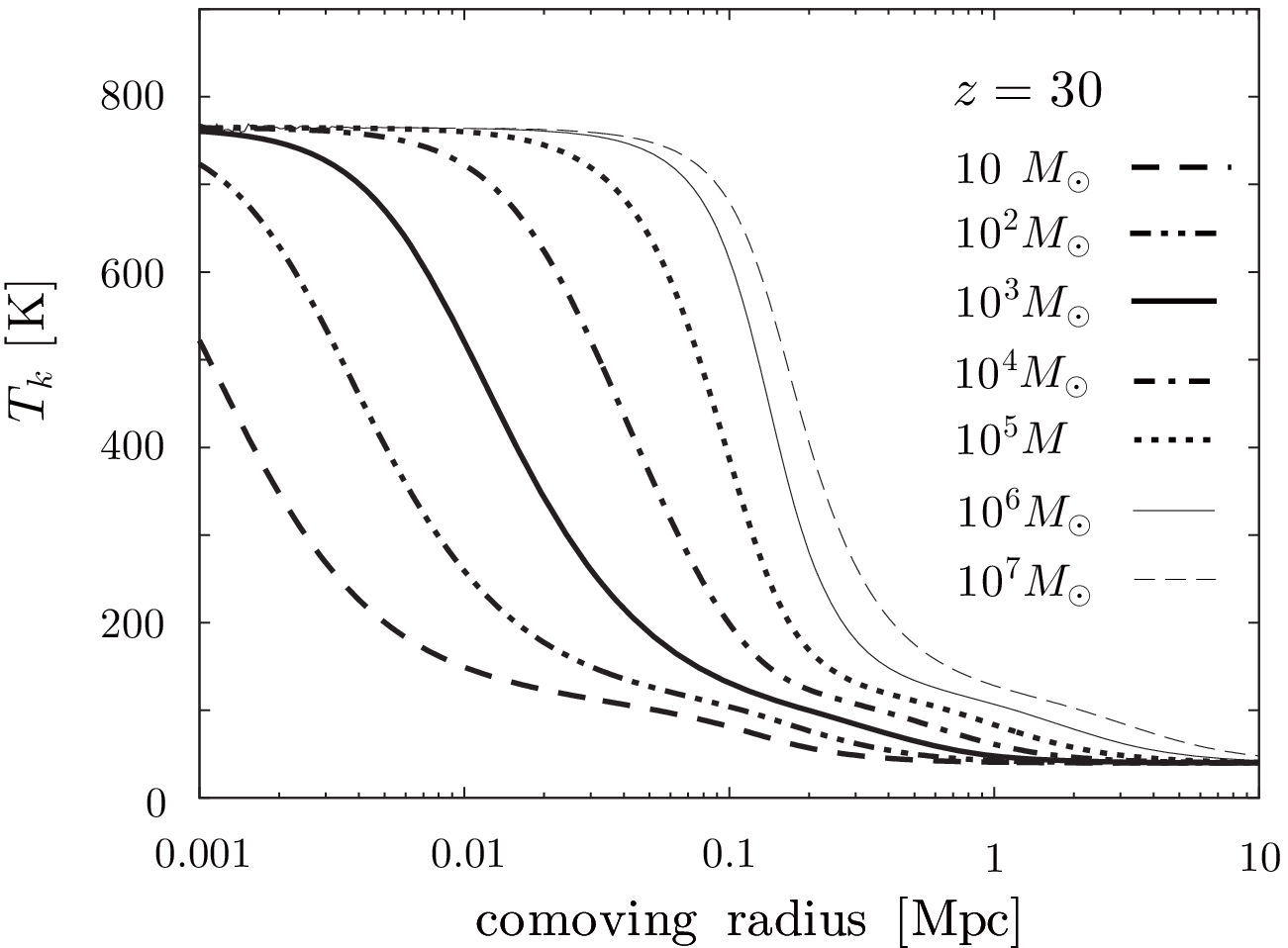}
  \end{center}
 \end{minipage}
 \begin{minipage}{0.5\hsize}
  \begin{center}
   \includegraphics[width=75mm]{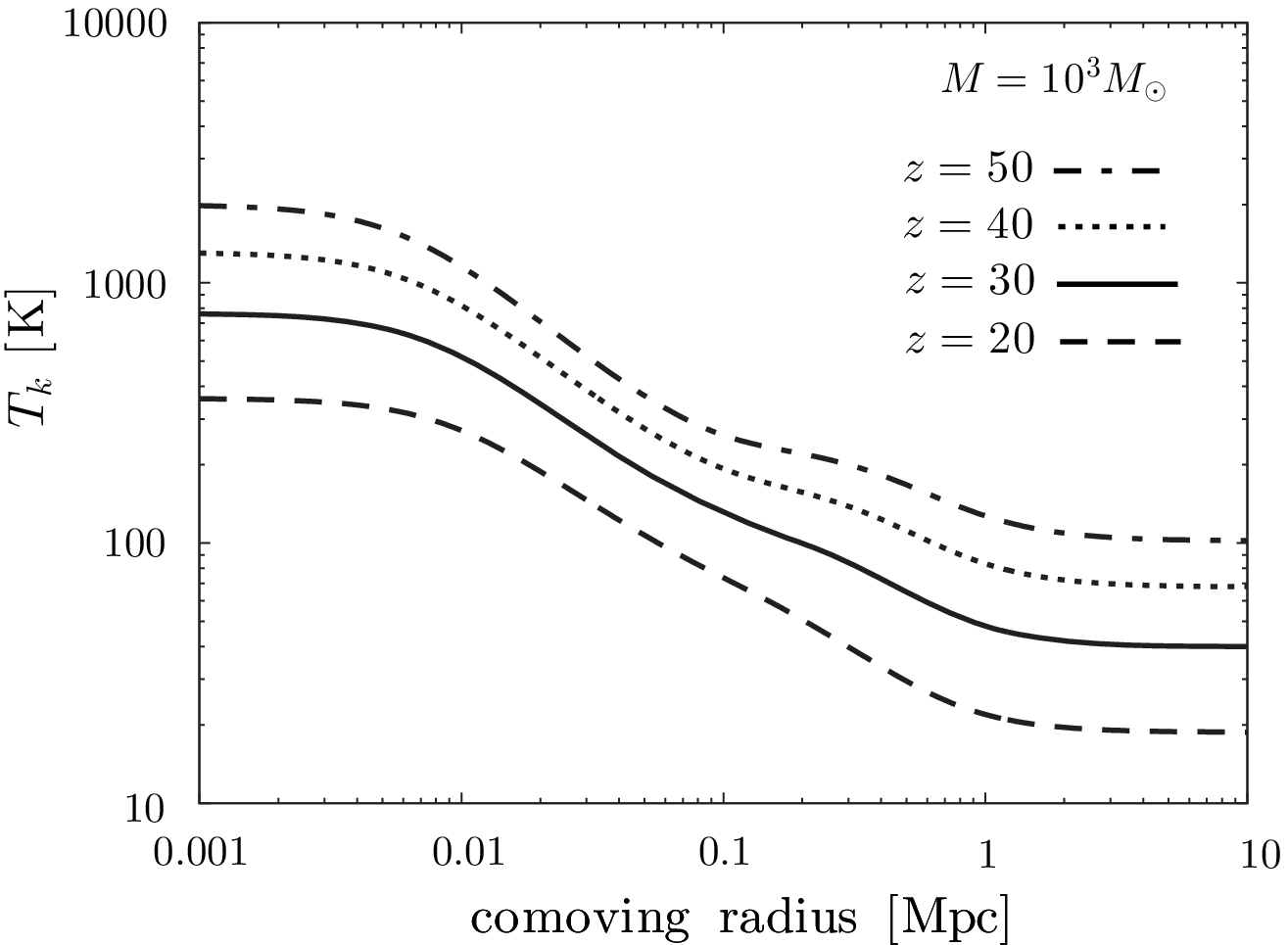}
  \end{center}
 \end{minipage}
\end{tabular}
\caption{The kinetic temperature of hydrogen gas as a function of the
 comoving radius. The left panel shows the dependence
 on PBH mass at $z=30$.
 The right panel shows the dependence on the redshift
 for a PBH with $M=10^3~M_\odot$.  In both panels, the representations of
 lines are same as in Fig.~\ref{fig:xh}.
}
\label{fig:TK}
\end{figure}

Fig.~\ref{fig:TK} shows the IGM kinetic temperature profiles for different
masses in the left panel and for different redshifts in the right
panel.   Here we add the background kinetic temperature to the kinetic
temperature in order to match both temperatures at a large distance
from a PBH.  
Near the
source, the temperature is determined by the heating rate and the
Compton cooling rate.
With increasing the distance from the source, the neutral fraction grows and
the optical depth $\tau$ becomes larger.
Accordingly, the number density of photons damps as shown in 
Eq.~(\ref{eq:flux_r}). As a result, the temperature starts to
decrease rapidly. Because the Compton cooling depends on the number of free electrons,
this cooling becomes ineffective at a distance where the neutral
fraction of hydrogen becomes almost unity.    For example, this scale
 corresponds to 0.1 comoving Mpc
for a PBH with $M=10^3~M_\odot$ at $z=30$.  
Beyond this point, the temperature mildly decreases  due to the cooling of the 
cosmic expansion. 
The kinetic temperature at the inner side is independent on a PBH
mass. However the region of the high temperature becomes larger as the PBH
mass increases.
In the right panel of Fig.~\ref{fig:TK}, the redshift dependence of the
kinetic temperature is shown.  
The larger the neutral hydrogen density is, the larger 
the heating efficiency becomes as in Eq.~(\ref{eq:heat}). Therefore the
temperature becomes high as the redshift increases.  

In this section, we obtained the neutral fraction of hydrogen by solving the equation of
the ionization-recombination equilibrium, Eq.~(\ref{eq:ionization_eq}).
However the recombination time scale is much smaller than the
ionization time scale at the redshifts we are interested in. Therefore
Eq.~(\ref{eq:ionization_eq}) might be incorrect.
In this limit, we can 
neglect the recombination effect. Without the
recombination term, we can write the evolution of the neutral fraction as
\begin{equation}
 {d x_{H} \over dt} = - \Gamma x_H.
\end{equation}
The solution of this equation is roughly $x_H \approx \exp (- \Gamma
/H)$.
Although the radial distance where $x_H $ reaches $0.8$ is same as in
the case of the ionization-recombination equilibrium, the resultant
ionization profile has a sharper edge. 
This is because most of the ionization photons are absorbed to ionize the inner
region. Accordingly, there are not enough ionization photons to ionize the outer region.
This fact also means that the heating efficiency is suppressed as the
radial distance increases, compared with the case of the ionization-recombination equilibrium.
As a result, the heated region becomes one-fifth times smaller than in
the previous case.

\section{21~cm Brightness temperature due to a PBH}

As shown in the previous section, a PBH ionizes and heats the surrounding region
by the X-ray emissions.   These ionization and heating leave an
observable signature as the differential
brightness temperature of the 21~cm intensity relative to the CMB
temperature.

The 21~cm intensity depends on the ratio between the number density of neutral
hydrogen in the excited state and in the ground state of the hyperfine structure.
This ratio can be quantified by the spin temperature $T_s$. 
The spin temperature is determined by the balance between three
processes: absorption of CMB photons, collisional excitation in the kinetic
temperature $T_k$, and Lyman $\alpha$ pumping \citep{1952AJ.....57R..31W,1958field}.
In the steady state approximation between these processes, the spin
temperature is obtained by
\citep{1958field}
\begin{equation}
T_{s} = \frac{T_\ast+T_{\gamma} +y_{k} T_{k} + y_\alpha
T_{k}}{1+y_{k}+y_\alpha},
\label{eq:t_spin}
\end{equation}
where
$T_\ast$ is 0.068 K which is the temperature
corresponding to the energy difference between the levels in the
hyperfine structure,
and
$y_{k}$ and $y_\alpha$
are the kinetic and Lyman-$\alpha$ coupling efficiencies, respectively.
The kinetic efficiency is given by
\begin{equation}
y_{k}=\frac{T_\ast}{A_{10} T_{k}}\left(C_H+C_e+C_p \right),
\end{equation}
where
 $A_{10}$ is the Einstein spontaneous emission rate coefficient,
$A_{10}=2.9 \times 10^{-15}~\rm{s^{-1}}$.
The terms, $C_H$, $C_e$ and
$C_p$, represent the de-excitation rates due to neutral hydrogen, electrons
and protons, respectively. Here we use the fitting formula by
\citet{2006ApJ...637L...1K},
\begin{equation}
C_H=n_H \kappa,
\quad
C_e=n_e \gamma_e,
\quad
C_p=3.2 n_p \kappa,
\label{eq:coefficientC}
\end{equation}
where
$n_e$ and $n_p$ are the electron and proton number
densities, respectively.  In Eq.~(\ref{eq:coefficientC}),
$\kappa$ is the effective single-atom rate coefficient,
\begin{equation}
\kappa = 3.1\times 10^{-11} n_H T_{k}^{0.357}
\exp(-32/T_{k})~ {\rm  cm^{3} s^{-1}},
\end{equation}
and $\gamma_e$ is given by 
$\log(\gamma_e/1~\rm{cm^3~s^{-1}}) = -9.607
+
0.5~\log~T_{k} \times \exp\left( -(\log~T_{k})^{4.5}/1800\right)$
for $T_{k} < 10^4~$K and $\gamma_e = \gamma_e(T_k=10^4~\rm K)$
for $T_k >10^4~$K.

The Lyman-$\alpha$ coupling efficiency is given by \citet{1958field},
\begin{equation}
y_\alpha= \frac{16 \pi^2 T_\ast e^2 f_{12} J_0}{27 A_{10} T_{k} m_e
c},
\label{eq:yalpha}
\end{equation}
where $f_{12}=0.416$ is the oscillator strength of the Lyman $\alpha$
transition, and $e$ and $m_e$ are the electron charge and
mass, respectively.
In Eq.~(\ref{eq:yalpha}), $J_0$ is the flux of the Lyman $\alpha$
photons due to collisional excitations. At the distance $r$ from the
source, $J_0$ can be written as \citep{2007MNRAS.375.1269Z} 
\begin{equation}
{J_0}(r) = \frac{\phi_\alpha\,c}{4 \pi H(z) \nu_\alpha } n_{H}
x_{H}(r)\int_{E_0}^{\infty}\sigma(E) {\cal N}(E;r) \frac{dE}{h \nu_\alpha},
\label{eq:excite}
\end{equation}
where $\phi_\alpha $ is the fraction of the absorbed energy going
into the collisional excitation of Lyman $\alpha$. \citet{1985ApJ...298..268S}
gave the following analytical form,
\begin{equation}
 \phi_\alpha \approx 0.48 \left( 1- (1-x_{H})^{0.27}\right)^{1.52}.
\end{equation}

\begin{figure}
 \begin{tabular}{cc}
 \begin{minipage}{0.5\hsize}
  \begin{center}
   \includegraphics[width=75mm]{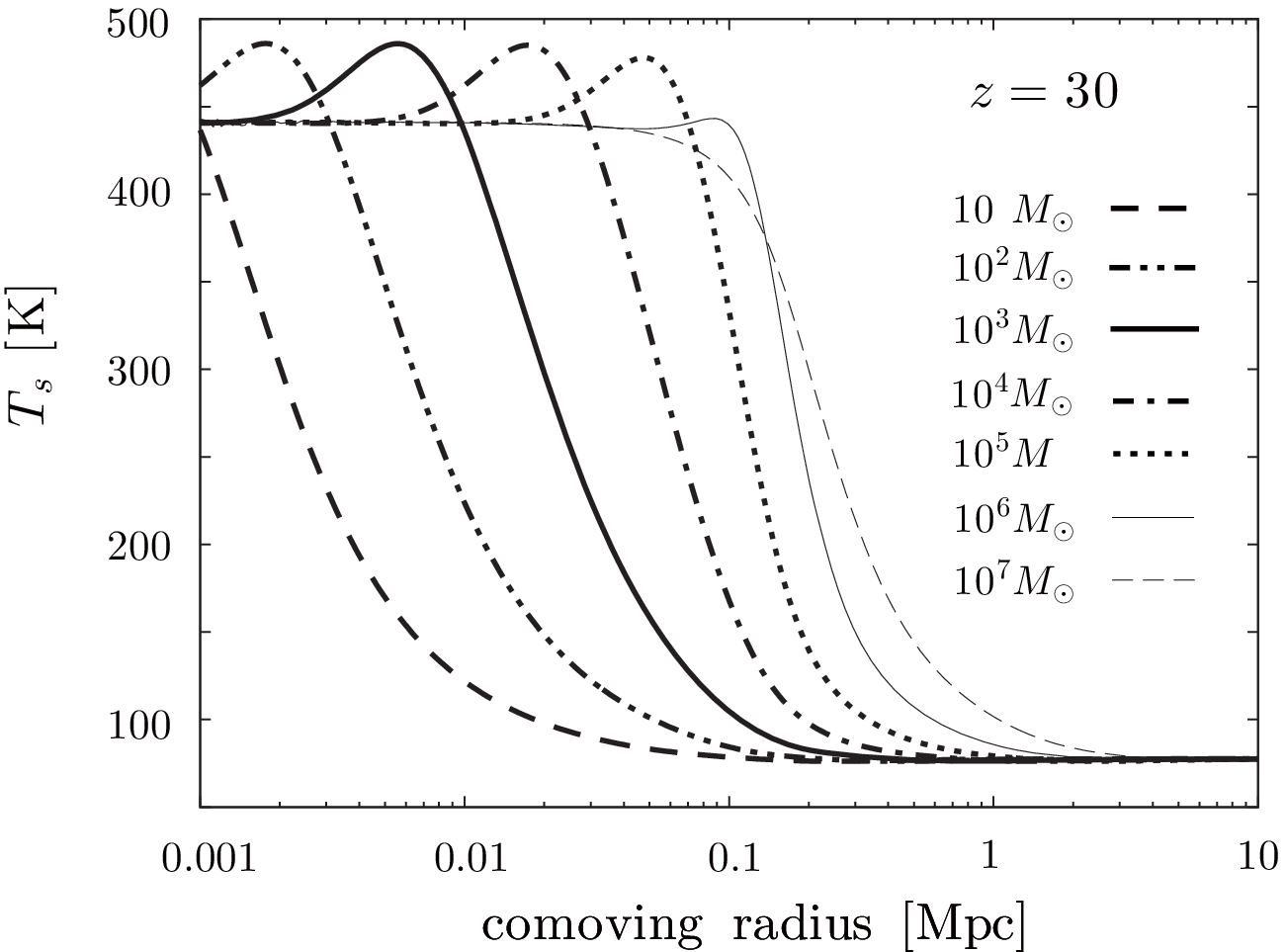}
  \end{center}
 \end{minipage}
 \begin{minipage}{0.5\hsize}
  \begin{center}
   \includegraphics[width=75mm]{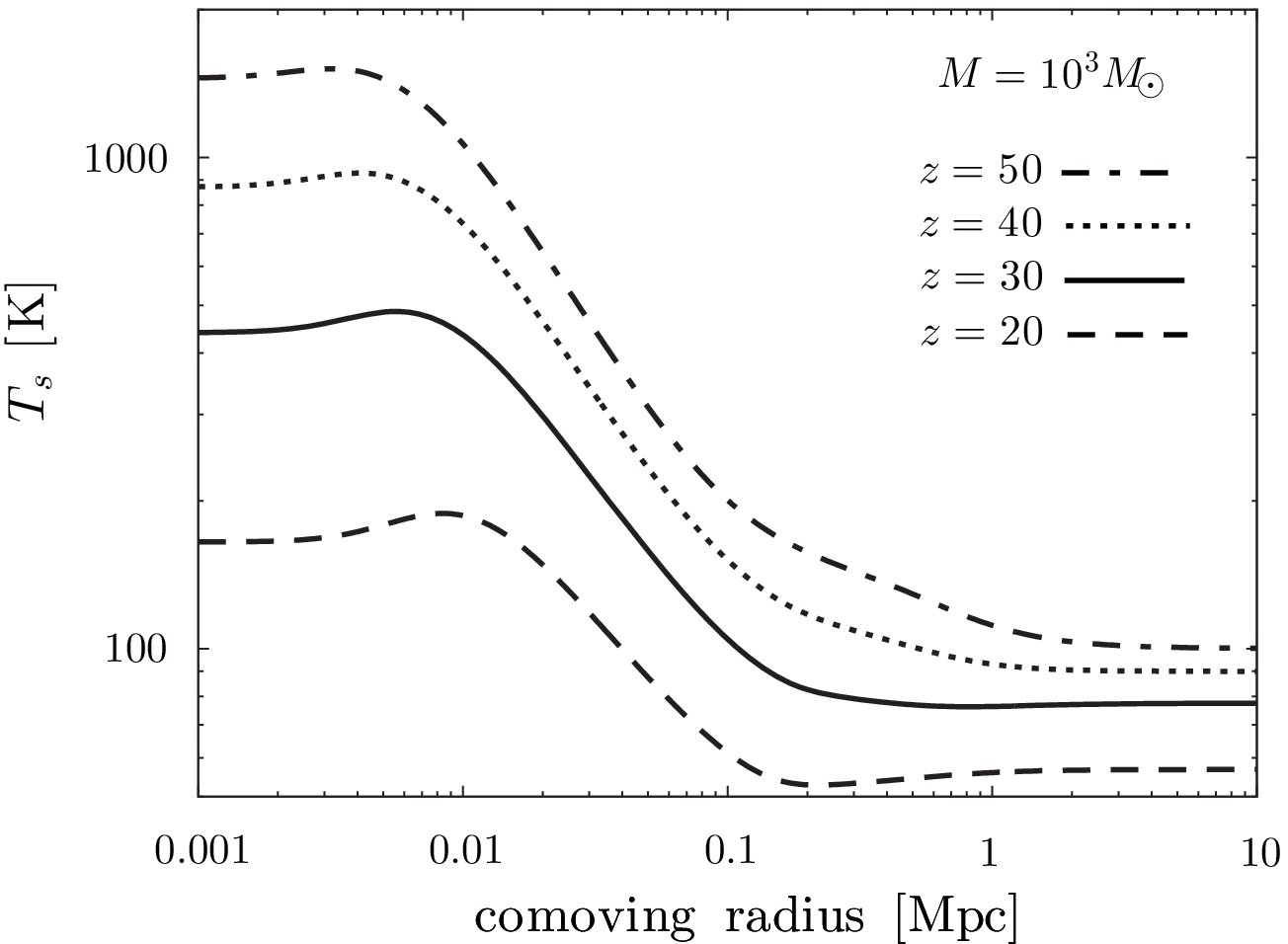}
  \end{center}
 \end{minipage}
\end{tabular}
\caption{The spin temperature as a function of the
 comoving radius.
The left panel shows the dependence
 on PBH mass at $z=30$.
 The right panel shows the dependence on the redshift
 for a PBH with $M=10^3~M_\odot$.
  In both panels, the representations of
 lines are same as in Fig.~\ref{fig:xh}.
}
\label{fig:Ts}
\end{figure}

Now we can calculate the spin temperature for a PBH
according to the results in the
previous section.  We show the results in Fig.~\ref{fig:Ts}. The left
panel shows the spin temperature for different masses of a PBH at $z=30$
and the right panel represents the spin temperature for different
redshifts for $M=10^3~M_\odot$.

In the highly ionized region, the profile of the spin temperature is flat
and does not depend on the PBH mass as shown in the left panel of
Fig.~\ref{fig:Ts}. Because the PBH can ionize surrouding gas,
the spin temperature is determined by
the CMB temperature and the term of the kinetic efficiency.
As the neutral fraction goes up with increasing distance, however,
the Lyman $\alpha$ efficiency $y_\alpha$ becomes larger so that 
the Lyman $\alpha$ coupling begins to be effective.
The peaks of $T_s$ at small radius for small PBH masses shown in the
left panel are caused due to this effect.  It should note that 
the peak values are also independent of mass since $T_s$ approaches
$T_k$ for a large value of $y_\alpha$. 
Because the flux of
Lyman-$\alpha$ quickly decreases as the radius increases, the spin temperature also
drops. Finally, the spin temperature settles down to the one
determined by the background CMB and kinetic temperatures.
The larger the PBH mass is, the larger 
the region where the spin temperature is higher than
the background one becomes.  

Because the kinetic temperature strongly depends on the redshift as
shown in the right panel of Fig.~\ref{fig:TK}, the amplitude of the spin
temperature also has strong dependence on the redshift as shown in the
right panel of Fig.~\ref{fig:Ts}. The spin temperature is almost same as
the kinetic temperature in the ionized region at high redshifts
($z>40$), since the kinetic efficiency $y_k$ dominates other
contributions in Eq.~(\ref{eq:t_spin}). In low redshifts ($z<30$), we
find the region where the spin temperature is lower than the background
one at the edge of the heated region. In these redshifts the background
spin temperature is almost the CMB temperature, while the background
spin temperature is between the CMB and kinetic temperatures in
redshifts, $30 < z < 50$. Since the kinetic temperature is lower than
the CMB temperature at the edge, the Lyman-$\alpha$ coupling is strong
and draws the spin temperature toward the kinetic temperature.
Therefore, there exists a region where the spin temperature becomes
lower than the background one. This tendency appears even in
\citet{2007MNRAS.375.1269Z}.

\begin{figure}
 \begin{tabular}{cc}
 \begin{minipage}{0.5\hsize}
  \begin{center}
   \includegraphics[width=75mm]{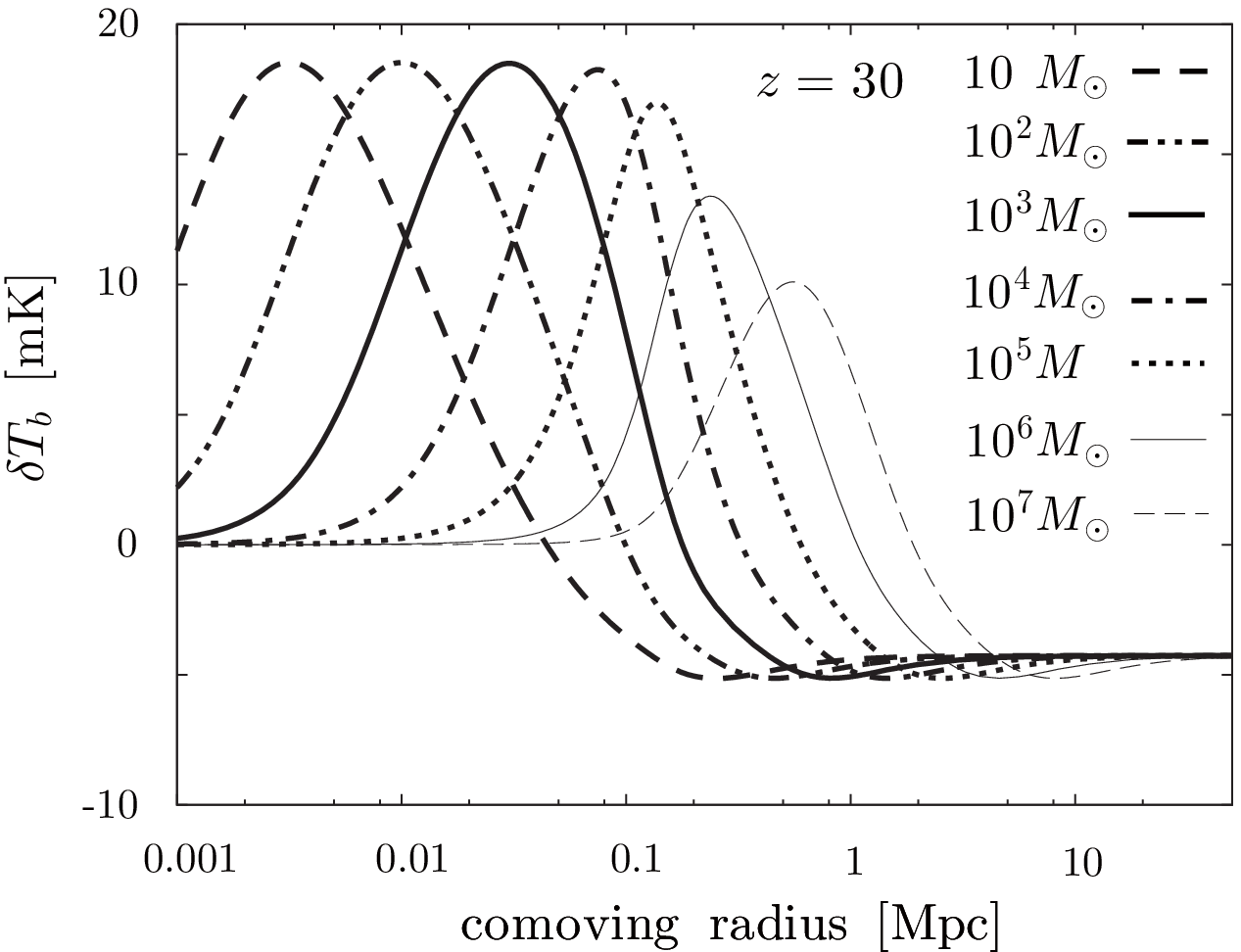}
  \end{center}
 \end{minipage}
 \begin{minipage}{0.5\hsize}
  \begin{center}
   \includegraphics[width=75mm]{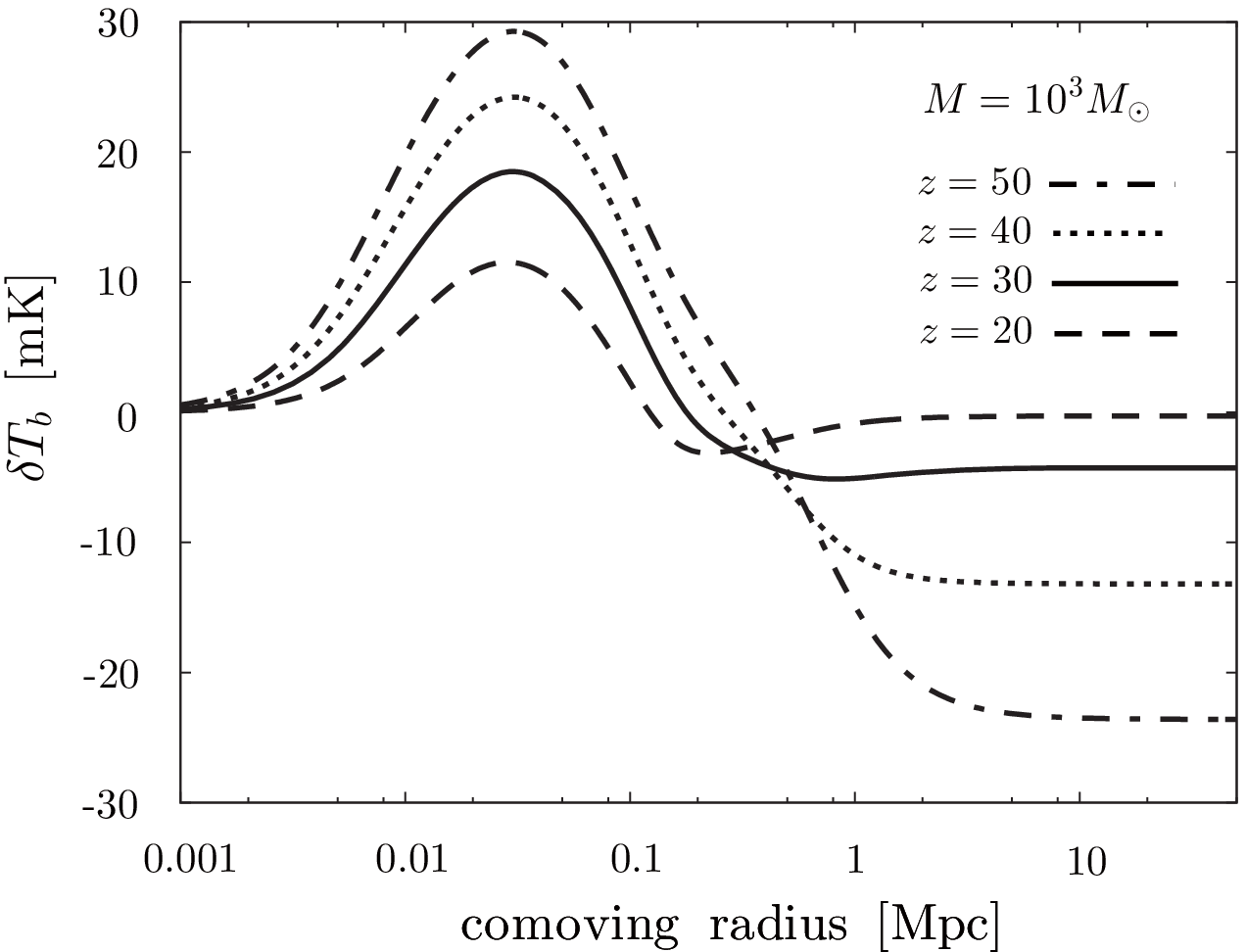}
  \end{center}
 \end{minipage}
\end{tabular}
\caption{The brightness temperature  fluctuations as a function of the
 comoving radius. The left panel shows the dependence
 on PBH mass at $z=30$. The right panel shows the dependence on the redshift
 for a PBH with $M=10^3~M_\odot$.  In both panels, the representations of
 lines are same as in Fig.~\ref{fig:xh}.
}
\label{fig:Tb}
\end{figure}

The differential brightness temperature from the CMB temperature for a given spin temperature 
$T_s$ is obtained by \citep{2003ApJ...596....1C}
\begin{equation}
\delta T_b =
\left(20~\mathrm{mK}\right)\left(1+\delta\right)\left(\frac{x_{H}}{h}\right)\left(1-\frac{T_{\gamma}}{T_{s}}\right)
\left(\frac{\Omega_{\rm B} h^2}{0.0223}\right)
\left[\left(\frac{1 +
z}{10}\right)\left(\frac{0.24}{\Omega_{\rm M}}\right)\right]^{1/2},
\label{eq:delta_tb}
\end{equation}
where $\delta$ is the density contrast.  Here  we assume $\delta = 0$ as
mentioned in section 2.  
In Fig.~\ref{fig:Tb},  $\delta T_b $ is shown as  a function of the comoving radial
distance from a PBH. The left panel is for different masses of a
PBH at $z=30$ and the right panel is for different redshifts for
$M=10^3~M_\odot$.  The $\delta T_b $  is almost zero near the source,
because hydrogen in such regions are totally ionized.
As the neutral fraction increases, the
brightness temperature also grows.
The peak amplitudes of $\delta T_b$ are independent on
the PBH mass for 
$M< 10^{4} ~M_\odot$.  This is because the spin temperature is much larger
than the CMB temperature, $T_s \gg T_\gamma$.  Accordingly,
from Eq.~(\ref{eq:delta_tb}),  
$\delta T_b$ becomes independent on $T_s$ and only depends on
$(1+z)^{1/2}$.    This redshift dependence is shown in the
right panel of Fig.~\ref{fig:Tb}.  
On the other hand, 
the peak amplitude becomes smaller for
a mass of the  PBH higher than $10^5 ~M_\odot$ in our model.

As in the case of $T_s$, the region
where the differential brightness temperature is below the background value exists
near the edge at $z<30$.    Eventually, however, $\delta T_b$ matches the
background value at a large distance.     
The size of the region where we can detect $\delta T_b$ 
becomes larger with increasing mass, and reaches almost 10 Mpc .

\section{The angular power spectrum of 21~cm brightness temperature due to PBHs}

In order to study the potential of cosmological 21~cm observations to
give a constraint on the PBH abundance, we evaluate the angular power
spectrum of 21~cm fluctuations due to PBHs with mass $M$.

The observed $\delta T_{\rm obs}$ at the direction $\hat {\bm n}$ is
the contribution from the IGM along the line of sight.
We can obtain $\delta T_{\rm obs}$ at the observation frequency $\nu$ 
by the integration over the comoving distance $r$.
\begin{equation}
 \delta T_{\rm obs} (\hat {\bm n}, \nu)
  = \int d r ~W(r ,\nu)  \delta T_b (r \hat{\bm n})
\label{eq:bright}
\end{equation}
where $W(r, \nu)$ is the
window function of a 21~cm observation with the frequency $\nu$ and
$\delta T_b({\bm x})$ is the differential brightness temperature at the
position ${\bm x}$, which is produced by PBHs with mass $M$.

In order to calculate the angular power spectrum, we employ the flat sky
approximation and the Limber
approximation, since we are only interested in the brightness temperature fluctuations on 
small scales ($\ell \gg 10$).
In these approximation, the angular power spectrum can be written in terms
of the three dimensional power spectrum of the brightness temperature \citep{Dodelson:2003ft}, 
\begin{equation}
 C_\ell (z_{\rm obs}) = \int dr \frac{W^2(r,\nu)  
  }{r^2} P(k, r),
  \label{eq:21cmcl}
\end{equation}
where $z_{\rm obs}$ is the redshift by choosing the
observation frequency $\nu = (1+z_{\rm obs}) \nu_{21}$ with
$\nu_{21} =1420.~$MHz, and
$P(k, r)$ is the power spectrum of the brightness temperature
fluctuations due to PBH with mass $M$ at $r$. 

We obtain the power spectrum of the brightness temperature fluctuations
$P(k, r)$, following the halo formalism \citep{2000MNRAS.318..203S}.
The power spectrum $P(k,r)$ can be divided to two contributions; the
Poisson contribution and clustering contribution.
On large scales, since PBHs may be biased tracers of the linear matter power
spectrum, they can make the clustering contribution. However, there is theoretical
uncertainty in PBH bias. To avoid such uncertainty,
we focus on only the Poisson contribution.
Therefore, the power spectrum due to PBHs is expressed
by
\begin{equation}
 P(k,r) = n_{\rm PBH} (M) |\Delta T_b^2 (k)|^2,
\label{eq:powerspectrum}
\end{equation}
where $n_{\rm PBH}(M)$ is the comoving number density of PBHs with $M$.

The comoving number density of PBHs is described by using the density
parameter $\Omega_{\rm PBH}$ as 
\begin{equation}
n_{\rm PBH} (M)= {\rho_c \Omega_{\rm PBH} \over  M} = 1.36 \times 10^{-2}
\left( {\Omega_{\rm PBH} \over 10^{-9}}     \right) \left( {M \over
 10^4 M_\odot} \right)^{-1} ~\rm Mpc^{-3},
 \end{equation}
where $\rho_c$ is the critical density
at the present epoch.
In Eq.~(\ref{eq:powerspectrum}), $\Delta T_b(k)$ is the Fourier transform of the brightness
temperature fluctuations from the background value,
\begin{equation}
 \Delta T_b(k) = 4 \pi \int x^2 dx (\delta T_b(x )-\delta T_{b0}) \frac{\sin(xk)}{xk},
\end{equation}
where $\delta T_b(x )$ is the differential brightness
temperature at the comoving radius $x$ from a PBH and $\delta T_{b0}$ is the background
brightness temperature given by Eqs.~(\ref{eq:t_spin}) and (\ref{eq:delta_tb})
with the background kinetic temperature, $T_{k0}$.

\begin{figure}
 \begin{tabular}{cc}
 \begin{minipage}{0.5\hsize}
  \begin{center}
   \includegraphics[width=75mm]{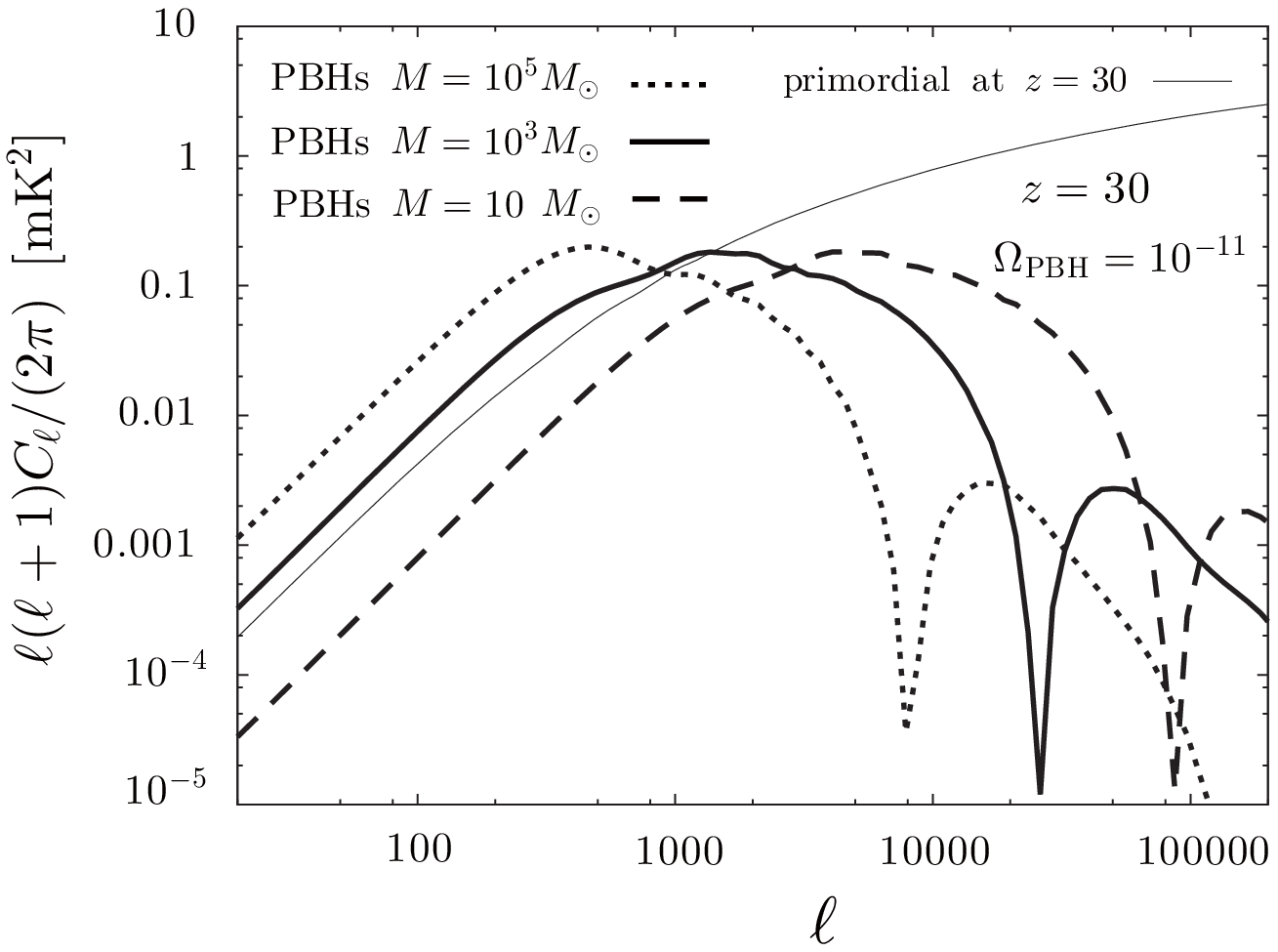}
  \end{center}
 \end{minipage}
 \begin{minipage}{0.5\hsize}
  \begin{center}
   \includegraphics[width=75mm]{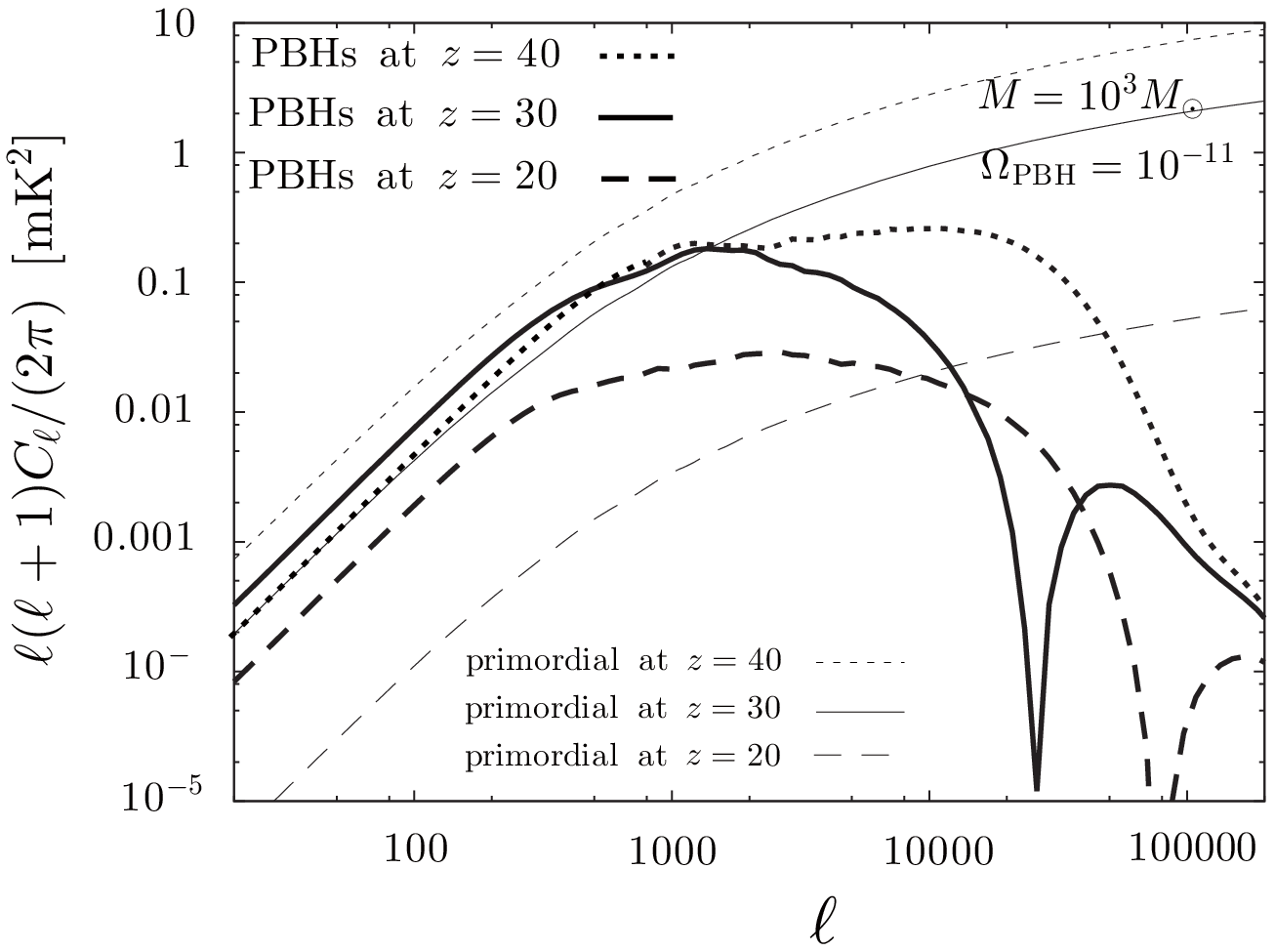}
  \end{center}
 \end{minipage}
\end{tabular}
\caption{The angular power spectrum of 21~cm brightness temperature
 fluctuations produced by PBHs as the function of
 comoving radius. The left panel shows the dependence
 on the PBH mass at $z=30$.
The dotted, solid and dashed lines represent for PBHs with $M=10^5 ~
 M_\odot$, $10^3~M_\odot$ and $10~M_\odot$, respectively.
 The right panel shows the dependence on the redshift
 for PBHs with $M=10^3~M_\odot$.
The dotted, solid and dashed lines are for PBHs at $z=40$, $z=30$ and
 $z=20$, respectively.
 We assume $\Omega_{\rm PBH} =
 10^{-11}$ in both panels. For comparison, we also plot the angular
 power spectrum due to the primordial density fluctuations as thin lines.
}
\label{fig:cl}
\end{figure}

Using $\delta T_b (x)$ obtained in the previous section as shown in
Fig.~\ref{fig:Tb},
we calculate the angular
power spectrum.  The results are  shown in  Fig.~\ref{fig:cl}.
Here we set
$\Omega_{\rm PBH} = 10^{-11}$.  For simplicity, we assume that the
window function has a strong peak at $r=r_{\rm obs}$ where $r_{\rm obs}$
corresponds to the comoving radial distance at $z_{\rm obs}$.
We approximate the window function as $W^2(r, r_{\rm obs}) =\delta(r-r_{\rm obs})$.
Since the promising signal of the cosmological 21~cm fluctuations is one due to the
primordial density fluctuations, we also plot the
angular power spectrum of those fluctuations obtained through CAMB \citep{2007PhRvD..76h3005L}.

In the left panel of Fig.~\ref{fig:cl},  it is shown that
the peak amplitude of the angular spectrum is independent on PBH's mass.
The peak locations shift toward large scales for massive PBHs. 
Moreover,  the power law index of the spectrum matches the one due to 
the primordial density fluctuations on  small
$\ell$'s.    The overall amplitude of the spectrum is simply
proportional to  $\Omega_{\rm PBH}$ or the PBH number
density.    For $\Omega_{\rm PBH}=10^{-11}$,   
it is shown that the angular spectrum of PBHs with
mass $10^3M_\odot$ matches  with the one due
to the primordial density fluctuations at $z=30$.  
It is clear that
the spectrum with heavier (lighter) PBHs  becomes comparable with
primordial density fluctuations' one if
$\Omega_{\rm PBH}$  is smaller (larger).
To match these two spectra, we find the critical value of $\Omega_{\rm PBH}$
as $\Omega_{{\rm PBH}c} \equiv 10^{-11} (M/10^{3}~ M_\odot)^{-0.2}$
for $z=30$.  
The PBH spectrum dominates over the one due to
the primordial density fluctuations if $\Omega_{\rm PBH}$ exceeds  
$\Omega_{{\rm PBH}c}$.  

The right panel of Fig.~\ref{fig:cl} shows that, although the peak
location does not depend on the redshift, the redshift dependence of the
peak amplitude is different from that due to the primordial density fluctuations.

As the result, while the PBH
contribution for $\Omega_{{\rm PBH}c}$ is subdominant at $z=40$,
even the PBHs with roughly $0.1 \times  \Omega_{{\rm PBH} c}$ can
produce the spectrum comparable with that due to the primordial density
fluctuations at $z=20$.

The brightness temperature profile at lower redshifts, $z<30$, has both
positive and negative peaks as shown in Fig.~\ref{fig:Tb}.  The reason
to have a negative peak is because there exists a region where the
brightness temperature is lower than the background, which is rather
difficult to see in the figure.  Accordingly, the angular power spectrum
due to PBHs also has two peaks at $z<30$ as shown in Fig.~\ref{fig:cl}.
The peak on a larger scale is due to the negative peak of $\Delta T_b$,
while that on a smaller scale is caused by the positive one.

At the last of this section, we discuss the constraint by a future
observation such as the square kilometer array (SKA)-like interferometer.
Measuring 21~cm anisotropies at high redshifts is a major challenge.
The sky at corresponding frequencies is contaminated by foreground due to synchrotron emissions from
the Galaxy and extragalactic sources.
The noise power spectrum of an observation including the beam effects is given by
\citep{1995PhRvD..52.4307K}
\begin{equation}
\frac{\ell(\ell +1)}{ 2 \pi}{ N_\ell ^{21}}= 
 { \ell (\ell+1) \over t_{\rm obs} \Delta \nu} \left(
{ D \lambda T_{\rm sys}\over A_{\rm eff}} 
\right)^2 \exp \left[
 {\frac{\ell(\ell+1)}{\ell_b^2}} \right] ,
\end{equation}
where $A_{\rm eff}$ is the effective area 
$T_{\rm sys}$ is the system temperature,
$t_{\rm obs}$ is the observation time, $\Delta \nu$ is
the frequency bandwidth, $D$ is the length of the baseline and
$\ell_b$ is given by $\ell _b = 4 \sqrt{\ln 2} /\theta_{fw}$ with
the resolution $\theta _{fw} \sim \lambda /D$.
Here we adopt the current design of
SKA\footnote{http://www.skatelescope.org/}; $ A_{\rm eff} =1~ {\rm km}^2$, $t_{\rm obs} = 1000~$hour, $\Delta \nu = 0.1~$MHz
and $D= 5~$km.
We set the system temperature to the sky temperature in minimum emission regions at high
Galactic latitudes given by $T_{\rm sys} = 180(\nu/180~{\rm
MHz})^{-2.6}~\rm K$.
In this design, the noise power spectrum is evaluated as
\begin{equation}
\frac{\ell(\ell +1)}{ 2 \pi}{ N_\ell ^{21}} \sim
0.9\times10^5 ~{\rm mK^2}~\left(\frac{\ell}{1000}\right)^2
\left(\frac{1+z}{30}\right)^{7.2}
\exp \left[
\left( \frac{\ell}{2500}  \frac{z+1}{30}\right)^2  \right].
\end{equation}
The noise power spectrum exponentially grows on larger multipoles than
$\ell_b  = 2500 (30/(1+z))$.
The amplitude of the angular power spectrum due to the PBHs is scaled by $\Omega_{\rm PBH}$.
Accordingly we find that, in order to dominate over the noise spectrum,
$\Omega_{\rm PBH} \sim 10^{-5}
(M/10^{3}~ M_\odot)^{-0.2}$ is required for $M>10 M_\odot$ at $z=30$ and
$\Omega_{\rm PBH} \sim 10^{-7} (M/10^{3}~ M_\odot)^{-0.2}$ is for $M>10 M_\odot$ at $z=20$.
Since the angular spectrum due to PBHs with mass less than
$10~M_{\odot}$ has a peak on
larger multipoles than $\ell_b$ in the SKA design,
there is no opportunity to measure the anisotropy spectrum 
due to such small mass PBHs by SKA.

\section{conclusion}

We have investigated the 21~cm signal produced by massive PBHs whose
masses are larger than $10~M_{\odot}$. Assuming a power-law spectrum 
of X-ray photons from an accretion disk, we have studied the ionization and heating of IGM gas
near  a PBH and evaluated the differential 21~cm brightness temperature.
We have shown that a PBH can induce an observable signal of differential 21 ~cm brightness
temperature.
The size of the region where we can find the differential brightness
temperature typically reaches 1--10~Mpc for our
interested PBH mass range. The exact size depends on the PBH mass.

We have also calculated the angular power spectrum of 21~cm fluctuations due
to PBHs. The peak position of the angular spectrum depends on the PBH mass,
while the amplitude is independent of the mass. Comparing this spectrum with the angular power
spectrum caused by primordial density fluctuations,  we have found that
both of them become comparable if 
$\Omega_{{\rm PBH}} =10^{-11} (M/10^{3}~
M_\odot)^{-0.2}$ at $z=30$ and $10^{-12} (M/10^{3}~
M_\odot)^{-0.2}$ at $z=20$ for PBH's mass from $10 ~ M_\odot $  to
$10^8~ M_\odot $.  If the density parameter is larger than these values,
the angular power spectrum due to PBHs exceeds the one from primordial
fluctuations and can be measured.
In other words, we cannot set constraints on the PBH density parameter below these
values from 21~cm observations. 
If we consider the sensitivity of the SKA-like
observation,  for example, we can detect the signal of PBHs up to 
$\Omega_{\rm PBH}=10^{-5} (M/10^{3}~
M_\odot)^{-0.2}$ at $z=30$ and $10^{-7} (M/10^{3}~
M_\odot)^{-0.2}$ at $z=20$ for PBHs with mass from $10^2 ~ M_\odot $  to
$10^8~ M_\odot $.

The ionization of IGM due to PBHs with such density parameters
does not affect the global reionization history of the universe since
reionization from each PBH only covers a tiny patch of the universe.  
Unlike reionization from first stars, therefore, such reionization has
little impact on CMB temperature anisotropies.
On the other hand, PBHs can heat IGM regions whose scale reaches 1-10Mpc
and 21~cm fluctuations are sensitive to the heated IGM regions.
Accordingly the PBH density parameter constrained
from WMAP data, that is $\Omega_{{\rm PBH}} < 10^{-7}$ 
\citep{2008ApJ...680..829R}, is several order of magnitude larger
than the values at which 21~cm fluctuations from PBHs comparable with
those from primordial density fluctuations as we mentioned above.    
In other words,  we can conclude that, if observation instruments or
foreground removal are more improved 
than the current SKA design, 21~cm fluctuation
observations have a potential to probe the PBH abundance which is
impossible to access by CMB observations.

The most theoretical uncertainty in this model is 
the flux of photons due to the accretion to PBHs.
In this paper, we assume that the X-ray photon flux amplitude is a tenth
of the Eddington luminosity and has the power law spectrum with $E^{-1}$ for simplicity.
We also studied the effect of the power law index by changing to $E^{-1/2}$.
The temperature profile shifts to a smaller scale due to the
decrease of the ionization efficiency. However this shift is small,
and the resultant angular power spectrum does not change much.

The amplitude of the luminosity is considered to depend on the matter accretion rate onto a
PBH.  \citet{2008ApJ...680..829R} have studied the luminosity for the 
Bondi-Hoyle accretion in detail. Although the luminosity depends on the
PBH mass and feedback effect on the ionization and temperature, they
have shown that the luminosity for a PBH with $M=10^3~ M_\odot$ is
roughly a hundredth of the Eddington luminosity at $z>20$.
In our model, 
the brightness temperature profile near a PBH at a certain
redshift depends only on the PBH's mass and
the amplitude of the angular power spectrum is scaled by $\Omega_{\rm
PBH}$.
Therefore, if we assume that the amplitude of the X-ray photon
flux is a hundredth of the Eddington luminosity, 
the required density parameter for PBHs to dominate the 21~cm
fluctuations due to primordial density fluctuations is 
$10^{-11} (M/10^{4}~
M_\odot)^{-0.2}$ at $z=30$ and $10^{-12} (M/10^{4}~
M_\odot)^{-0.2}$ at $z=20$ for PBH's mass from $10 ~ M_\odot $  to
$10^8~ M_\odot $.

%
%

\section*{Acknowledgments}

We thank A. Long for useful comments.
H. T. is supported by the DOE at Arizona state university.
N. S. is supported by
Grand-in-Aid for Scientific Research No. 22340056.  
This research has also been supported in part by
World Premier International Research Center Initiative,
MEXT, Japan.

 
\end{document}